\documentclass[a4paper,11pt,reqno,oldfontcommands]{article}
\usepackage{amsmath}
\usepackage{graphicx}
\usepackage{amssymb}
\usepackage{float}
\usepackage{authblk}
\usepackage{siunitx}
\usepackage{hyperref}
\title{Laser-induced forced evaporative cooling of molecular anions below 4 Kelvin}

\author[1]{Jonas Tauch}
\author[1]{Saba Z. Hassan}
\author[2]{Markus N\"otzold}
\author[2]{Eric S. Endres}
\author[2]{Roland Wester}
\author[1]{Matthias Weidem\"uller}

\affil[1]{Physikalisches Institut, Ruprecht-Karls-Universit\"at Heidelberg, Im Neuenheimer Feld 226, 69120 Heidelberg, Germany}
\affil[2]{Institut f\"ur Ionenphysik und Angewandte Physik, Universit\"at Innsbruck, Technikerstrasse 25, 6020 Innsbruck, Austria}

\date{}

\begin{document}
	
\maketitle

%\singlespacing
\textbf{The study of cold and controlled molecular ions is pivotal for fundamental research in modern physics and chemistry \cite{Tomza2019}. Investigations into cooling molecular anions, in particular, have proven to be of key consequence for the production of cold antihydrogen \cite{Kellerbauer2006}, the creation and study of anionic Coulomb crystals \cite{Bonitz2008,Drewsen2003} as well as in atmospheric research \cite{Vuitton2009} and astrochemistry \cite{McCarthy2006,Larsson2012,Millar2017a}. The commonly used anion cooling technique via collisions with a buffer gas is limited by the temperature of the used cryogenic cooling medium. Here, we demonstrate forced evaporative cooling of anions via a laser beam with photon energies far above the photodetachment threshold of the anion \cite{Cru1990,Keller2019}. We reach runaway evaporative cooling of an anionic OH$^{-}$ ensemble from an initial temperature of 370(12)\;Kelvin down to 2.2(8)\;Kelvin. This corresponds to a three orders of magnitude increase in the ions' phase space density approaching the near-strong Coulomb coupling regime. A quantitative analysis of the experimental results, via a full thermodynamic model including the role of intrinsic collisional heating, represents the anion cooling dynamics without any fitting parameters. This technique can be used to cool, in principle, any anionic species below liquid helium temperature, providing a novel tool to push the frontiers of anion cooling to much lower than the state-of-the-art temperature regimes.}
%\section*{Introduction}
\newpage
Recent technological developments in the hybrid trapping of ions and atoms have provided key insights into the interaction dynamics of molecular ions with atoms \cite{Tomza2019}. These studies involve investigations of the atom-ion reactive collisions \cite{Puri2017,Hall2012,Germann2014,Kilaj2018,Haerter2013,Joschka2017,Puri2019}, as well as the elastic and inelastic collisions, cooling the ions in their external (translational) and internal (vibration and rotational) degrees of freedom respectively \cite{Hudson2016,Willitsch2012,Rellergert2013,Stoecklin2016,Hauser2015}. 
While most of these systems, primarily involve cationic ions, cooling anions to the cold and even ultracold regime is of great interest in the study of astrochemically relevant reaction dynamics, i.e. the growth of anionic carbon chains detected in Titan's atmosphere \cite{McCarthy2006,Herbst1981,Bastian2019}, associative detachment reactions of hydrogen in the early universe \cite{Kreckel2010,Gerlich2012} or proton exchange formation pathways of ammonia in interstellar media \cite{Otto2008,Gianturco2019}.
The outermost electron of a negative ion is bound by a short-range potential, which results in a lack of multiple stable electronic states, and due to their fragile nature, most anions cannot be cooled via traditional laser cooling methods. Although few promising candidates exist \cite{Keller2015,Gerber2018,Wang2017}, they still need precooling techniques, and direct Doppler cooling of anions has been, so far, experimentally elusive. 

Building upon the success of evaporative cooling to create ultracold atoms, forced evaporative cooling of anions was proposed three decades ago \cite{Cru1990}. For a static laser beam position, laser-induced evaporative cooling was demonstrated in atomic anions, where O$^-$ anions were cooled from 1.15\;eV to 0.33\;eV \cite{Keller2019}. Alternatively, evaporative cooling of trapped antiprotons to temperatures as low as 9\;\si{\kelvin} was achieved by altering the trap's potential landscape \cite{Andresen2010}. In our work, we employ this approach of laser-induced evaporative cooling to molecular anions and thus demonstrate the generality of this method. We reach temperatures below 4\;\si{\kelvin}, which sets the limit of buffer-gas cooling using cryogenic helium. We remove the high-energetic tail of a molecular anionic ensemble, stored in a radio-frequency (rf) ion trap, via selective photodetachment. Subsequent rethermalization leads to efficient cooling of the remaining ions, which can be further enhanced by sweeping the detachment laser into the ion cloud ("forced evaporative cooling").
%By introducing a dynamically moving photon flux, the high-energy ions are eliminated, resulting in the translational cooling of the ions via laser-induced forced evaporative cooling.
The cooling process is well modeled by an ab-initio thermodynamic calculation, revealing the role of radio-frequency heating induced by ion-ion collisions. With this approach, we cool down an ensemble of OH$^{-}$ anions from 370(12)\;K to 2.2(8)\;K. We achieve a Coulomb coupling parameter of $\Gamma=0.06(5)$ in less than 4 seconds along with an increase in the ensemble's relative phase space density by three orders of magnitude. 
%\section*{Main text}
\begin{figure}[H]
\makebox[\textwidth]{\includegraphics[width=183mm]{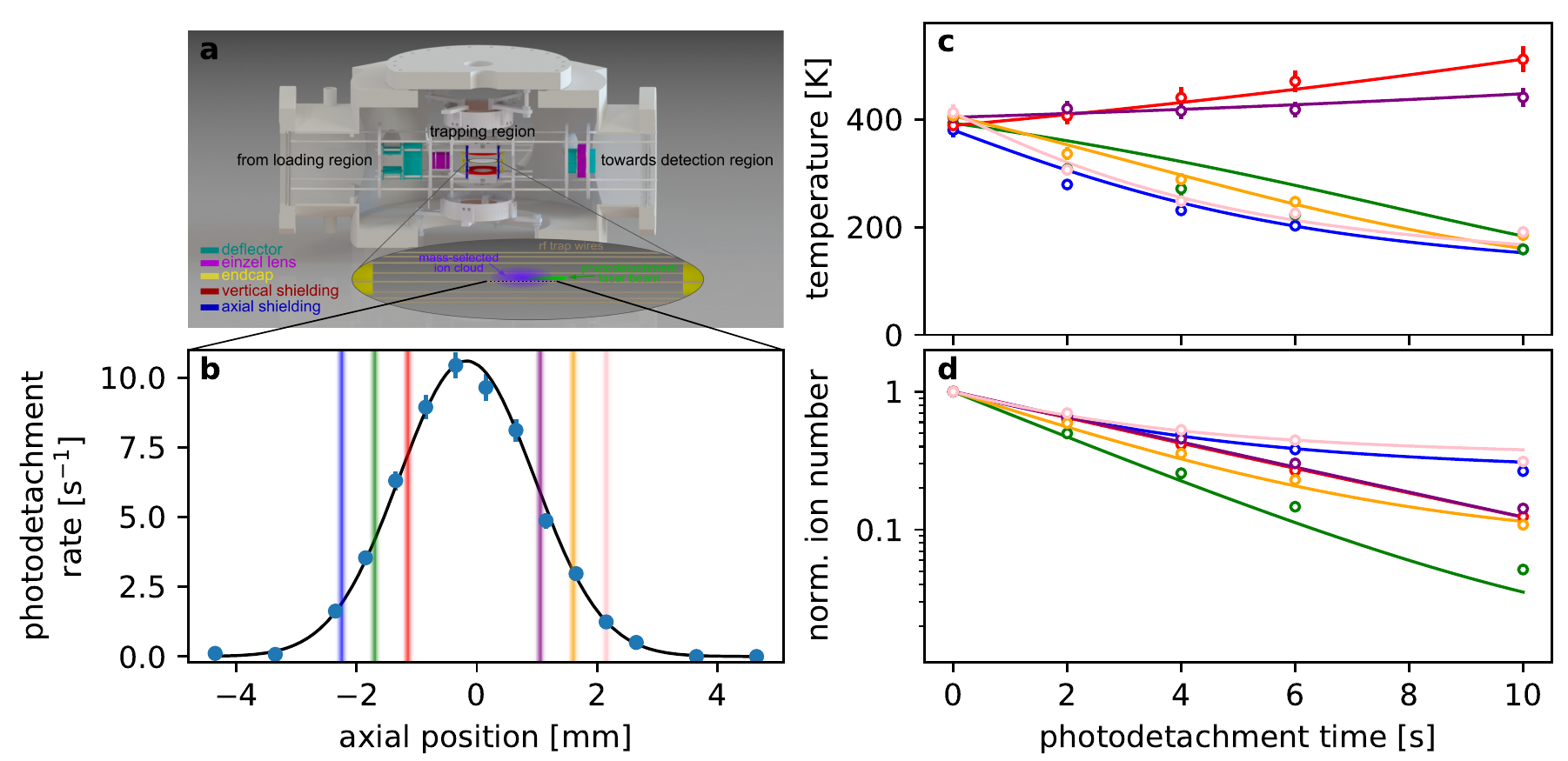}}
\caption{\textbf{Experimental setup and thermalization dynamics under position-dependent photodetachment.} \textbf{(a)}, Schematic representation of the octupole wire trap with hollow-core endcaps for loading and extraction of the ions. Perpendicular to the trap axis, we shine in a laser beam for photodetachment tomography and energy selective evaporation. \textbf{(b)}, Axial ion distribution in the trap at 370(12)\;K. The solid points represent photodetachment rates, which are proportional to the local column density. Each rate is determined by fitting an exponential decay to a loss measurement depending on the photodetachment time. The error bars show the $\pm 1\sigma$ uncertainty of these fits. The solid curve is a fitted Gaussian function. Six different laser beam positions and their width are illustrated by colored vertical lines. The right panels show \textbf{(c)} the temporal evolution of temperature and \textbf{(d)} the normalized ion number for the six laser beam positions. The open circles represent the measured data and the solid lines show the numerical integration of Equation \ref{eq:dT} and \ref{eq:dN}.}
\label{fig1}
\end{figure}
In our experiment, we trap a cloud of OH$^-$ anions in a radio-frequency octupole wire trap (see Figure \ref{fig1}~\textbf{a}). A description of the experimental details are given in the Methods section. By addressing ions in a chosen axial interval, i.e. via photodetachment with a focused laser beam we can manipulate the ensemble's mean energy. To perform evaporative cooling, ions with total energy higher than the mean energy must be removed \cite{Cru1990}.  With this approach, one is able to preserve the trapping potential landscape, compared to the forced evaporative cooling of ions by lowering the trap depth \cite{Andresen2010}.
We measured the evolution of temperature and the number of ions for six laser beam positions. 
The position and width of the laser beam compared to the ions' initial distribution for the six measurements are illustrated in Figure \ref{fig1}~\textbf{b} as colored vertical bars. 

The evolution of ensemble temperature and number of ions is measured as a function of the interaction time with the photodetachment light, as shown in Figure \ref{fig1}~\textbf{c}, \ref{fig1}~\textbf{d} respectively. For photodetachment laser beam position, sufficiently far away from the center of the ion cloud, we observe a reduction in the ion temperature by a factor of 2 within ten seconds of interaction time with the laser beam. If the photodetachment beam is closer to the center, the ions undergo heating. During this interaction, the number of ions are reduced by roughly a factor of 10.

Based on ref. \cite{Cru1990}, the change in temperature (T) and ion number (N) in an octupole trap is described by:
\begin{align}
\dot{T}=&\frac{3}{7}\Bigg(-\frac{\sigma_\text{pd}}{k_\text{B}}\int\rho_\text{ion}(x,y,z,T)\Phi(x,z,x_\text{L},z_\text{L})V(x,y,z)dxdydz+\label{eq:dT}\\
&\frac{5\sigma_\text{pd}}{6}T\int\rho_\text{ion}(x,y,z,T)\Phi(x,z,x_\text{L},z_\text{L})dxdydz \Bigg),\nonumber \\
\dot{N}=&-\sigma_\text{pd}N\int\rho_{ion}(x,y,z,T)\Phi(x,z,x_\text{L},z_\text{L})dxdydz.\label{eq:dN}
\end{align}
$\sigma_\text{pd}$ is the absolute photodetachment cross section ($\sigma_\text{OH$^-$}=8.5\cdot10^{-22}$\,m$^2$ at 660 nm \cite{Hlavenka2009a}), $\rho_\text{ion}$ the normalized ion distribution and $\Phi(x,z,x_\text{L},z_\text{L})$ the Gaussian distributed photon flux at the position $(x_\text{L},z_\text{L})$. The total potential $V$ is defined as the sum of the harmonic axial potential created by the static endcaps and the time averaged rf potential in radial direction. In an octupole trap this ponderomotive potential is proportional to $r^6$. The first term of Equation \ref{eq:dT} corresponds to the mean potential energy of the ions addressed by the photodetachment beam and the second term represents the mean potential energy of the entire ensemble. The prefactors $\frac{3}{7}$ and $\frac{5}{6}$ stem from applying the virial theorem. A detailed description of the model can be found in the Methods section. For evaporative cooling, the ensemble's thermalization rate must be fast as compared to the induced loss rate. To ensure the validity of this condition the laser power is adjusted accordingly for different laser beam positions (see Supplementary Information, Section I). Thus, as long as the ion-ion thermalization rate surpasses the ion loss rate, the process of evaporative cooling (and heating) can be described via this equilibrium picture. 

Equations \ref{eq:dT} and \ref{eq:dN} are integrated numerically using the measured experimental parameters without any free parameter. The results of this integration are shown in Figures \ref{fig1}~\textbf{c} and \textbf{d}, respectively, as solid lines for each laser beam configuration. It can be seen that the model accurately describes the dynamics. The ensemble is cooled for sufficiently large distances of the laser beam from the cloud center reaching temperatures around 200\,K after 10\,s, while removing about 90\% of the atoms.
\begin{figure}[H]
	\noindent
	\makebox[\textwidth]{\includegraphics[width=183mm]{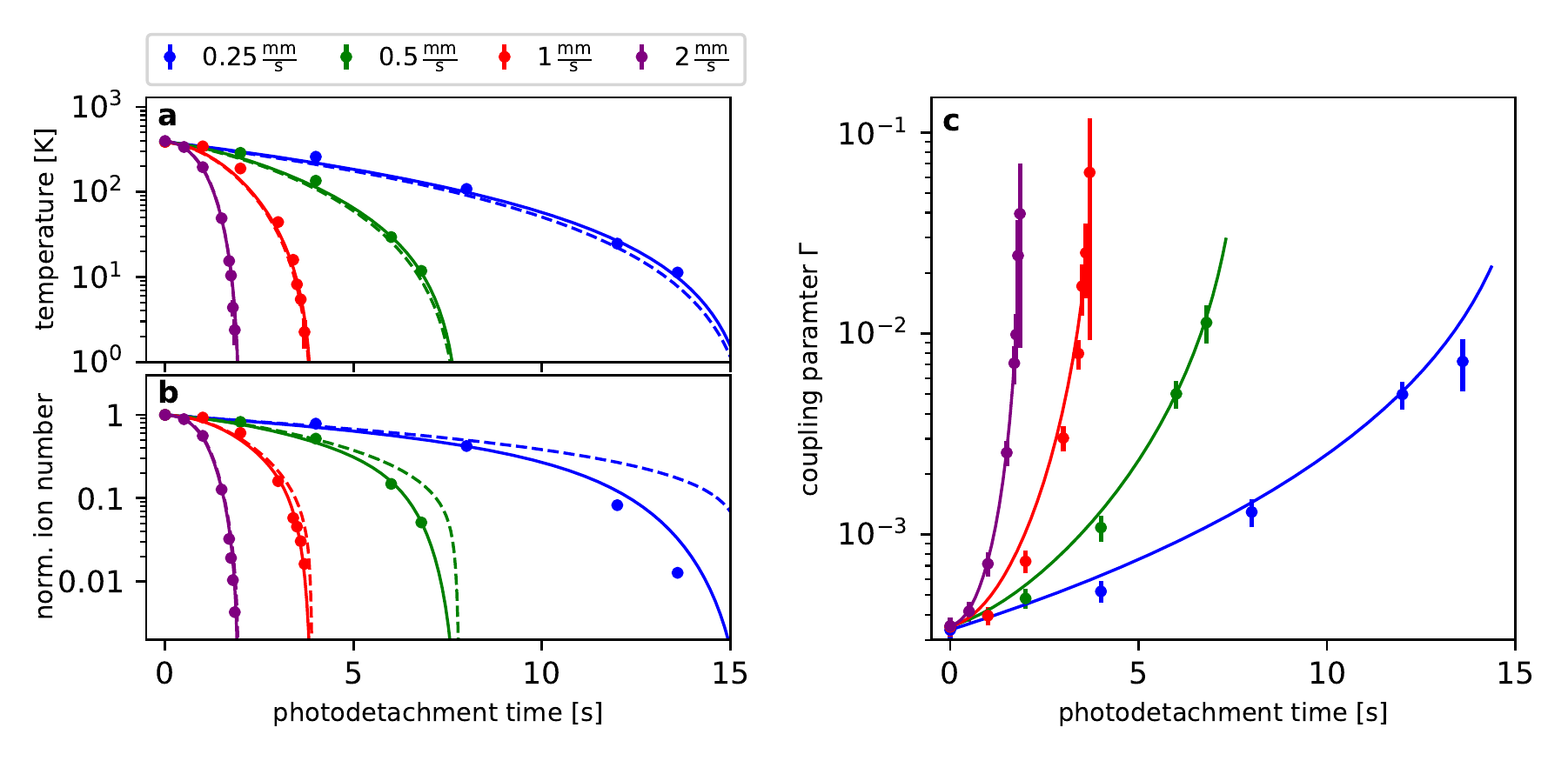}}
	\caption{\textbf{Forced evaporative cooling via photodetachment.} Evolution of (\textbf{a}) the ion ensemble's temperature, (\textbf{b}) normalized ion number and (\textbf{c}) the Coulomb coupling constant, while moving the laser beam towards the trap center for 4 different beam velocities. The initial beam position is 4\;mm off-center. For each beam velocity, ions are extracted at the positions 4\;mm, 3\;mm, 2\;mm, 1\;mm and 0.6\;mm from the center. For 1\;mm/s and 2\;mm/s additional measurements at 0.5\;mm, 0.4\;mm and 0.3\;mm where taken. The measured experimental results are shown as solid points. The dashed lines represent the result of the numerical integration of Equation \ref{eq:dT} and \ref{eq:dN}. The solid lines include the heating rate of Equation \ref{eq:heat}. (\textbf{c}) shows measured Coulomb coupling parameter as solid points and the model including the heating rate as solid lines. The model is terminated at the residual ion number of 1\;\% of the initial ion population.}
	\label{fig2}
\end{figure}

Enhanced cooling can be achieved by moving the laser dynamically towards the trap center.
%In the following experiment, we positioned the photodetachment laser beam away from the trap center. As the detachment light was switched on, the laser beam was simultaneously moved with a constant velocity towards the trap center.The change in the ion temperatures and ion numbers is measured as described in the Methods.
The resulting ion dynamics is shown for four different laser beam velocities in Figures \ref{fig2}~\textbf{a} and \textbf{b}.
Moving the laser beam sufficiently slowly to the trap center the final ensemble temperature is 2.2(8)\;K with about 2\;\% of the initially loaded ions remaining. This corresponds to an increase in phase space density by three orders of magnitude. 

The dashed lines shown in Figure \ref{fig2}~\textbf{a} and \textbf{b} are the numerical integration of Equation~\ref{eq:dT} and \ref{eq:dN} including a dynamically moving photodetachment laser beam. Although the thermodynamic model describes the observed temperature evolution very well, the measured number of ions deviates significantly from the predicted numbers for lower beam velocities. This can be explained by the influence of ion-ion collisions, which transfer the energy of the driving rf field to the ions' secular motion. The resulting heating rate can be approximated by the product of the ion-ion collision rate, described by the Chandrasekhar-Spitzer plasma self-collision rate $\nu_{ii}(T)$ \cite{Spitzer}, and the relative energy transfer per collision $\bar{\epsilon}$:
\begin{align}
\dot{T}=\frac{3}{7}\nu_\text{ii}(T)\bar{\epsilon}\frac{1}{3}T=\frac{3}{7}\frac{ne^4\log\Lambda_\text{ii}}{12\epsilon_0^2\sqrt{\pi^3m(k_\text{B}T)^3}}\bar{\epsilon}\frac{1}{3}T.\label{eq:heat}
\end{align}
The mean rf motion in an octupole trap is given by $\langle E_\text{rf}\rangle=\frac{1}{3}k_\text{B}T$ and the relative energy change per collision $\bar{\epsilon}=0.045(6)$ is measured by a background heating measurement (see Supplementary Information, Section II). $n$ is the ion density and $\log\Lambda_\text{ii}$ the Coulomb logarithm, with $\Lambda_\text{ii}$ being the ratio between the Debye length and the classical distance of the closest approach.

Equation \ref{eq:dT} and \ref{eq:dN} are independent on the absolute ion number, but the ion-ion collision rate is not. The detection efficiency of our setup is therefore determined by an independent thermalization measurement (see Supplementary Information, Section I). %With these two additional measurements, we retrieve both parameters, the detection efficiency and $\bar{\epsilon}$, of Equation \ref{eq:heat}. 
The zero fit-parameter models, including the heating rate, are shown as the solid lines in Figure \ref{fig2}. The complete model describes the change in both, the temperature and the ion number, very accurately. 

The efficiency and pace of cooling our ensemble are limited by two timescales, the evaporation rate and the thermalization rate. The evaporation rate is proportional to the detachment rate, and thus photon flux. If we move the laser beam faster than the removal of high-energy ions, the cooling efficiency drops, as can be seen in Figure \ref{fig2}. The thermalization rate determines how fast the energy is redistributed over the whole ensemble.
If we move the laser beam too fast, the evaporation rate exceeds the thermalization rate, and the ions are only spilled without an increase in the ensembles' phase space density. If we move it too slowly, rf heating via ion-ion collisions becomes more prominent and results in an additional ion loss. The thermalization rate of a one-component plasma is proportional to the product of the Coulomb logarithm and the ensembles' phase space density, for which we see a tremendous increase by three orders of magnitude (see Supplementary Information, Section III). 

%To look into the future prospects of this technique, it has to be ensured that the ion-ion collisional heating can be described by the Chandrasekhar-Spitzer self-collision rate.
A one-component plasma is characterized by the dimensionless Coulomb coupling constant $\Gamma$, which represents the nearest-neighbor Coulomb energy to the thermal energy:
\begin{align}
\Gamma=\frac{e^2}{4\pi\epsilon_0k_\text{B}T\langle r\rangle},
\label{eq:gamma}
\end{align}
where $\langle r\rangle$ is the mean inter-particle distance (Wigner-Seitz radius). In a weakly coupled plasma, $\Gamma\ll 1$, Coulomb interaction can be treated as two particle collisions perturbing the system, as we did in Equation \ref{eq:heat}. For $\Gamma ~ 1$, the assumption breaks down and reveals phenomena as liquid-like behavior and a liquid-solid phase transition at $\Gamma=178$ \cite{Ichimaru1986}\cite{Gilbert}. 
Figure \ref{fig2}~\textbf{c} shows the  Coulomb coupling constant corresponding to the measurement shown in the two left panels. For the maximum forced evaporative cooling efficiency (laser beam velocity of 1\;mm/s), the coupling parameter increases from initially $\Gamma=0.35(4)\cdot 10^{-3}$ to 0.06(5) resulting in runaway evaporative cooling. Based on the model, the coupling constant could be further increased by increasing the detachment rate beyond the rate currently achievable in our experimental setting (see Supplementary Information, Section IV).

%This implies that if adapt the photon flux to this tremendous increase in thermalization rate, we can optimize the runaway regime and attain considerably higher cooling efficiencies. Considering the same initial conditions as in our above-shown measurements, one reaches the runaway regime in less than 1\; second by keeping the relative cut-off energy constant.

%To ensure the validity of our model, the evaporation rate is chosen to be 10\;\% of the thermalization rate. A more detailed description of the runaway approach can be found in the .

%Our experimental results and the model show that with forced evaporative cooling via above-threshold photodetachment one can, in principle, attain temperatures much lower than cryogenic liquid helium, the current state-of-the-art for anion cooling.
Our experimental results and the model show that with forced evaporative cooling via above-threshold photodetachment is an efficient tool to reduce the thermal energy of, in principle, any anionic species by orders of magnitude. This establishes a way to translationally cool anions below the temperature of liquid helium without the need of a buffer gas, thus ensuring ultra-high vacuum conditions. This would allow to enter the regime of low-energy scattering resonances \cite{Lauderdale1983,Sanche1990,Engel2019}, as has recently been achieved with atomic cations \cite{Weckesser2021}, eventually approaching the regime of strong Coulomb coupling.
In a multi-species anionic plasma, this laser-induced cooling provides the advantage of addressing one anionic species, while sympathetically cooling the other species, e.g. antiprotons, without the loss of the latter.
The cooling of the molecular anions can even be extended to address internal degrees of freedom by tuning the laser light frequency close to the anion's photodetachment threshold \cite{Cru1990}.
\newpage
\section*{Methods}
\subsection*{Experimental details}
Our molecular anion of interest, the hydroxide OH$^{-}$ anion, is created in a pulsed plasma discharge source and subsequently trapped in an octupole wire trap. This trap combines the advantage of reduced rf heating in multipole traps and the use of wires to provide optical access to investigate atom-anion interactions in a hybrid trap \cite{Kas2019,Hassan2022a,Hassan2022b}. 
The measurements were performed on trapped ions initially thermalized with a room temperature helium buffer gas, which results in an ensemble with a temperature slightly above the buffer gas. This increase in temperature is attributed due to rf heating \cite{Thermometry}. We determine the temperature of ions in the trap by extracting them through a hollow-core endcap and subsequently detecting the time-of-flight (TOF) for each particle on a micro-channel plate (MCP). With an optimized extraction potential, one can linearly map the energy distribution onto the TOF to the detector. A detailed description of the setup and the TOF thermometry can be found in \cite{Thermometry}.

While the energy distribution is obtained from the TOF, the ions' spatial distribution in the trap can be measured via above-threshold photodetachment tomography \cite{Trippel2006}. In our setup the laser light with a photon energy of 1.88\;eV, which is above the photodetachment threshold of OH$^{-}$ (1.83\;eV \cite{Smith1997}), is focused down to a diameter of \SI{180}{\micro\meter} (1/e$^{2}$) and can be moved by a translational stage.
One measurement of the ion distribution along the trap axis, after initialization with the helium buffer gas pulse, is shown in Figure \ref{fig1}~\textbf{b} as solid points. Since the axial potential is well-described by a harmonic potential, a Gaussian distribution of the ion cloud can be observed.
\subsection*{Thermodynamic model}
\label{methods:model}
The approach of this rate model is based on \cite{Cru1990} and it assumes that the cooling process is slow enough, that the ions are at any time in thermal equilibrium. This way the total energy can be described by the sum of the mean kinetic energy and the mean potential energy:
\begin{align}
E_\text{tot}=\langle E_\text{k}\rangle_v+\langle E_\text{p} \rangle_r=\frac{3}{2}k_\text{B}T+\frac{5}{6}k_\text{B}T=\frac{7}{3}k_\text{B}T.
\end{align}
The prefactor $\frac{5}{6}$ is the result of the virial theorem for an octupole trap. The change in ion number is described by the overlap integral.
\begin{align}
\dot{N}=-N\langle\langle\sigma\Phi\rangle_r\rangle_v = - N \int \sigma(\textbf{v})\Phi(\textbf{r})\rho_\text{ion}(\textbf{r},\textbf{v})d\textbf{r}d\textbf{v}
\end{align}
$\sigma$ is the absolute photodetachment cross section, $\rho_\text{ion}$ the normalized ion density and $\Phi$ is the photon flux. The change of the total energy is given by the lost ions' energy:
\begin{align}
\dot{E}_\text{tot} &= -N\langle\langle\sigma\Phi E\rangle_r\rangle_v\\
\Rightarrow \langle\langle\dot{E}\rangle_r\rangle_v &= -\langle\langle\sigma\Phi E\rangle_r\rangle_v+\langle\langle\sigma\Phi\rangle_r\rangle_v\langle\langle E\rangle_r\rangle_v\\
&= -\langle\langle\sigma\Phi (E_\text{k}+E_\text{p})\rangle_r\rangle_v+\langle\langle\sigma\Phi\rangle_r\rangle_v\langle\langle (E_\text{k}+E_\text{p})\rangle_r\rangle_v
\end{align}
Only the laser intensity and potential energy depends on \textbf{r} and the kinetic energy on \textbf{v}. In the case of above-threshold photodetachment, the cross section is independent of the ions' velocity. Thus the kinetic energy terms cancel out.
\begin{align}
%\langle\langle\dot{E}\rangle_r\rangle_v =-&\sigma\langle\langle\rho_{ion}\PhiE_k\rangle_r\rangle_v-\sigma\langle\rho_{ion}\Phi E_p\rangle_r+\sigma\langle\rho_{ion}\Phi\rangle_r\langle E_k\rangle_v+\sigma\langle\rho_{ion}\Phi\rangle_r\langle E_p\rangle_r\\
\langle\langle\dot{E}\rangle_r\rangle_v =-&\sigma \langle\Phi E_\text{p}\rangle_r + \sigma \langle\Phi\rangle_r \langle E_\text{p}\rangle_r\\
\Rightarrow\frac{7}{3}k_\text{B}\dot{T} = -&\sigma\int\rho_\text{ion}(x,y,z,T)\Phi(x,z,x_\text{L},z_\text{L})V(x,y,z)dxdydz +\\
&\frac{5}{6}k_\text{B}T\sigma\int\rho_\text{ion}(x,y,z,T)\Phi(x,z,x_\text{L},z_\text{L})dxdydz \nonumber\\
\dot{N}=-&\sigma N\int\rho_\text{ion}(x,y,z,T)\Phi(x,z,x_\text{L},z_\text{L})dxdydz- k_\text{bgr}N
\end{align}
The background loss rate has been measured to be $k_\text{bgr}=0.009(1)$\;s$^{-1}$.
The potential is determined by photodetachment tomography measurements at 370(12)\;K in axial and radial direction assuming a quadratic and a $r^6$-Potential:
\begin{align}
V(x,y,z)=2.04(9) \cdot 10^{-15}\si{\J\per\meter^2}\cdot z^2+3.9(6)\cdot 10^{-3}\si{\J\per\meter^6}\cdot\left(x^2+y^2\right)^3
\end{align}

\newpage
\newpage
\bibliographystyle{fec}
\bibliography{Bibliography/Paper-ForcedEvaporativeCooling-FinalCitationList}

\begin{thebibliography}{10}
\newcommand{\enquote}[1]{``#1''}
\expandafter\ifx\csname url\endcsname\relax
  \def\url#1{\texttt{#1}}\fi
\expandafter\ifx\csname urlprefix\endcsname\relax\def\urlprefix{URL }\fi
\providecommand{\eprint}[2][]{\url{#2}}

\bibitem{Tomza2019}
M.~Tomza, K.~Jachymski, R.~Gerritsma, {et~al.}
\newblock \enquote{{Cold hybrid ion-atom systems}}.
\newblock \href{http://dx.doi.org/10.1103/RevModPhys.91.035001}{{Rev. Mod.
  Phys.}} \href{http://dx.doi.org/10.1103/RevModPhys.91.035001}{{\textbf{91},
  035001}}\href{http://dx.doi.org/10.1103/RevModPhys.91.035001}{{ (2019)}}.

\bibitem{Kellerbauer2006}
A.~Kellerbauer and J.~Walz.
\newblock \enquote{{A novel cooling scheme for antiprotons}}.
\newblock \href{http://dx.doi.org/10.1088/1367-2630/8/3/045}{{New J. Phys.}}
  \href{http://dx.doi.org/10.1088/1367-2630/8/3/045}{{\textbf{8}}}\href{http://dx.doi.org/10.1088/1367-2630/8/3/045}{{
  (2006)}}.

\bibitem{Bonitz2008}
M.~Bonitz, P.~Ludwig, H.~Baumgartner, {et~al.}
\newblock \enquote{{Classical and quantum Coulomb crystals}}.
\newblock \href{http://dx.doi.org/10.1063/1.2839297}{{Phys. Plasmas.}}
  \href{http://dx.doi.org/10.1063/1.2839297}{{\textbf{15}(5)}}\href{http://dx.doi.org/10.1063/1.2839297}{{
  (2008)}}.

\bibitem{Drewsen2003}
M.~Drewsen, I.~Jensen, J.~Lindballe, {et~al.}
\newblock \enquote{{Ion Coulomb crystals: A tool for studying ion processes}}.
\newblock \href{http://dx.doi.org/10.1016/S1387-3806(03)00259-8}{{Int. J. Mass
  Spectrom.}}
  \href{http://dx.doi.org/10.1016/S1387-3806(03)00259-8}{{\textbf{229}(1-2),
  83}}\href{http://dx.doi.org/10.1016/S1387-3806(03)00259-8}{{ (2003)}}.

\bibitem{Vuitton2009}
V.~Vuitton, P.~Lavvas, R.~V. Yelle, {et~al.}
\newblock \enquote{{Negative ion chemistry in Titan's upper atmosphere}}.
\newblock \href{http://dx.doi.org/10.1016/j.pss.2009.04.004}{{Planet. Space
  Sci.}} \href{http://dx.doi.org/10.1016/j.pss.2009.04.004}{{\textbf{57}(13),
  1558}}\href{http://dx.doi.org/10.1016/j.pss.2009.04.004}{{ (2009)}}.

\bibitem{McCarthy2006}
M.~C. McCarthy, C.~A. Gottlieb, H.~Gupta, and P.~Thaddeus.
\newblock \enquote{{ Laboratory and Astronomical Identification of the Negative
  Molecular Ion C$_6$H$^-$}}.
\newblock \href{http://dx.doi.org/10.1086/510238}{{ApJ}}
  \href{http://dx.doi.org/10.1086/510238}{{\textbf{652}(2),
  L141}}\href{http://dx.doi.org/10.1086/510238}{{ (2006)}}.

\bibitem{Larsson2012}
M.~Larsson, W.~D. Geppert, and G.~Nyman.
\newblock \enquote{{Ion chemistry in space}}.
\newblock \href{http://dx.doi.org/10.1088/0034-4885/75/6/066901}{{Rep. Prog.
  Phys.}}
  \href{http://dx.doi.org/10.1088/0034-4885/75/6/066901}{{\textbf{75}(6)}}\href{http://dx.doi.org/10.1088/0034-4885/75/6/066901}{{
  (2012)}}.

\bibitem{Millar2017a}
T.~J. Millar, C.~Walsh, and T.~A. Field.
\newblock \enquote{{Negative ions in space}}.
\newblock \href{http://dx.doi.org/10.1021/acs.chemrev.6b00480}{{Chem. Rev.}}
  \href{http://dx.doi.org/10.1021/acs.chemrev.6b00480}{{\textbf{117}(3),
  1765}}\href{http://dx.doi.org/10.1021/acs.chemrev.6b00480}{{ (2017)}}.

\bibitem{Cru1990}
A.~Crubellier.
\newblock \enquote{{Theory of laser evaporative cooling of trapped negative
  ions. I. Harmonically bound ions and RF traps}}.
\newblock \href{http://dx.doi.org/10.1088/0953-4075/23/20/020}{{J. Phys. B: At.
  Mol. Opt. Phys.}}
  \href{http://dx.doi.org/10.1088/0953-4075/23/20/020}{{\textbf{23}(20),
  3585}}\href{http://dx.doi.org/10.1088/0953-4075/23/20/020}{{ (1990)}}.

\bibitem{Keller2019}
G.~Cerchiari, P.~Yzombard, and A.~Kellerbauer.
\newblock \enquote{{Laser-Assisted Evaporative Cooling of Anions}}.
\newblock \href{http://dx.doi.org/10.1103/PhysRevLett.123.103201}{{Phys. Rev.
  Lett.}}
  \href{http://dx.doi.org/10.1103/PhysRevLett.123.103201}{{\textbf{123}(10),
  103201}}\href{http://dx.doi.org/10.1103/PhysRevLett.123.103201}{{ (2019)}}.

\bibitem{Puri2017}
P.~Puri, M.~Mills, C.~Schneider, {et~al.}
\newblock \enquote{{Synthesis of mixed hypermetallic oxide BaOCa$^+$ from
  laser-cooled reagents in an atom-ion hybrid trap}}.
\newblock \href{http://dx.doi.org/10.1126/science.aan4701}{{Science}}
  \href{http://dx.doi.org/10.1126/science.aan4701}{{\textbf{357}(6358),
  1370}}\href{http://dx.doi.org/10.1126/science.aan4701}{{ (2017)}}.

\bibitem{Hall2012}
F.~H.~J. Hall and S.~Willitsch.
\newblock \enquote{{Millikelvin Reactive Collisions between Sympathetically
  Cooled Molecular Ions and Laser-Cooled Atoms in an Ion-Atom Hybrid Trap}}.
\newblock \href{http://dx.doi.org/10.1103/PhysRevLett.109.233202}{{Phys. Rev.
  Lett.}}
  \href{http://dx.doi.org/10.1103/PhysRevLett.109.233202}{{\textbf{109},
  233202}}\href{http://dx.doi.org/10.1103/PhysRevLett.109.233202}{{ (2012)}}.

\bibitem{Germann2014}
M.~Germann, X.~Tong, and S.~Willitsch.
\newblock \enquote{{Observation of electric-dipole-forbidden infrared
  transitions in cold molecular ions}}.
\newblock \href{http://dx.doi.org/10.1038/nphys3085}{{Nat. Phys.}}
  \href{http://dx.doi.org/10.1038/nphys3085}{{\textbf{10}(11),
  820}}\href{http://dx.doi.org/10.1038/nphys3085}{{ (2014)}}.

\bibitem{Kilaj2018}
A.~Kilaj, H.~Gao, D.~R{\"o}sch, {et~al.}
\newblock \enquote{{Observation of different reactivities of para and
  ortho-water towards trapped diazenylium ions}}.
\newblock \href{http://dx.doi.org/10.1038/s41467-018-04483-3}{{Nat. Commun.}}
  \href{http://dx.doi.org/10.1038/s41467-018-04483-3}{{\textbf{9}(1),
  2096}}\href{http://dx.doi.org/10.1038/s41467-018-04483-3}{{ (2018)}}.

\bibitem{Haerter2013}
A.~H{\"a}rter, A.~Kr{\"u}kow, M.~Dei{\ss}, {et~al.}
\newblock \enquote{{Population distribution of product states following
  three-body recombination in an ultracold atomic gas}}.
\newblock \href{http://dx.doi.org/10.1038/nphys2661}{{Nat. Phys.}}
  \href{http://dx.doi.org/10.1038/nphys2661}{{\textbf{9}(8),
  512}}\href{http://dx.doi.org/10.1038/nphys2661}{{ (2013)}}.

\bibitem{Joschka2017}
J.~Wolf, M.~Deiß, A.~Krükow, {et~al.}
\newblock \enquote{{State-to-state chemistry for three-body recombination in an
  ultracold rubidium gas}}.
\newblock \href{http://dx.doi.org/10.1126/science.aan8721}{{Science}}
  \href{http://dx.doi.org/10.1126/science.aan8721}{{\textbf{358}(6365),
  921}}\href{http://dx.doi.org/10.1126/science.aan8721}{{ (2017)}}.

\bibitem{Puri2019}
P.~Puri, M.~Mills, I.~Simbotin, {et~al.}
\newblock \enquote{{Reaction blockading in a reaction between an excited atom
  and a charged molecule at low collision energy}}.
\newblock \href{http://dx.doi.org/10.1038/s41557-019-0264-3}{{Nat. Chem.}}
  \href{http://dx.doi.org/10.1038/s41557-019-0264-3}{{\textbf{11}(7),
  615}}\href{http://dx.doi.org/10.1038/s41557-019-0264-3}{{ (2019)}}.

\bibitem{Hudson2016}
E.~R. Hudson.
\newblock \enquote{{Sympathetic cooling of molecular ions with ultracold
  atoms}}.
\newblock \href{http://dx.doi.org/10.1140/epjti/s40485-016-0035-0}{{EPJ Techn.
  Instrum.}}
  \href{http://dx.doi.org/10.1140/epjti/s40485-016-0035-0}{{\textbf{3}(1),
  8}}\href{http://dx.doi.org/10.1140/epjti/s40485-016-0035-0}{{ (2016)}}.

\bibitem{Willitsch2012}
S.~Willitsch.
\newblock \enquote{{Coulomb-crystallised molecular ions in traps: methods,
  applications, prospects}}.
\newblock \href{http://dx.doi.org/10.1080/0144235X.2012.667221}{{Int. Rev.
  Phys. Chem.}}
  \href{http://dx.doi.org/10.1080/0144235X.2012.667221}{{\textbf{31}(2),
  175}}\href{http://dx.doi.org/10.1080/0144235X.2012.667221}{{ (2012)}}.

\bibitem{Rellergert2013}
W.~G. Rellergert, S.~T. Sullivan, S.~J. Schowalter, {et~al.}
\newblock \enquote{{Evidence for sympathetic vibrational cooling of
  translationally cold molecules}}.
\newblock \href{http://dx.doi.org/10.1038/nature11937}{{Nature}}
  \href{http://dx.doi.org/10.1038/nature11937}{{\textbf{495}(7442),
  490}}\href{http://dx.doi.org/10.1038/nature11937}{{ (2013)}}.

\bibitem{Stoecklin2016}
T.~Stoecklin, P.~Halvick, M.~A. Gannouni, {et~al.}
\newblock \enquote{{Explanation of efficient quenching of molecular ion
  vibrational motion by ultracold atoms}}.
\newblock \href{http://dx.doi.org/10.1038/ncomms11234}{{Nat. Commun.}}
  \href{http://dx.doi.org/10.1038/ncomms11234}{{\textbf{7}(1),
  11234}}\href{http://dx.doi.org/10.1038/ncomms11234}{{ (2016)}}.

\bibitem{Hauser2015}
D.~Hauser, S.~Lee, F.~Carelli, {et~al.}
\newblock \enquote{{Rotational state-changing cold collisions of hydroxyl ions
  with helium}}.
\newblock \href{http://dx.doi.org/10.1038/nphys3326}{{Nat. Phys.}}
  \href{http://dx.doi.org/10.1038/nphys3326}{{\textbf{11}(6),
  467}}\href{http://dx.doi.org/10.1038/nphys3326}{{ (2015)}}.

\bibitem{Herbst1981}
E.~Herbst.
\newblock \enquote{{Can negative molecular ions be detected in dense
  interstellar clouds?}}
\newblock \href{http://dx.doi.org/10.1038/289656a0}{{Nature}}
  \href{http://dx.doi.org/10.1038/289656a0}{{\textbf{289}(5799),
  656}}\href{http://dx.doi.org/10.1038/289656a0}{{ (1981)}}.

\bibitem{Bastian2019}
B.~Bastian, T.~Michaelsen, J.~Meyer, and R.~Wester.
\newblock \enquote{{Anionic Carbon Chain Growth in Reactions of
  C$_2^-$,C$_4^-$,C$_6^-$, C$_2$H$^-$ , C$_4$H$^-$ , and C$_6$H$^-$ with
  C$_2$H$_2$}}.
\newblock \href{http://dx.doi.org/10.3847/1538-4357/ab2042}{{ApJ}}
  \href{http://dx.doi.org/10.3847/1538-4357/ab2042}{{\textbf{878}(2),
  162}}\href{http://dx.doi.org/10.3847/1538-4357/ab2042}{{ (2019)}}.

\bibitem{Kreckel2010}
H.~Kreckel, H.~Bruhns, M.~{\v{C}}{\'{i}}{\v{z}}ek, {et~al.}
\newblock \enquote{{Experimental Results for H$_2$ Formation from H$^-$ and H
  and Implications for First Star Formation}}.
\newblock \href{http://dx.doi.org/10.1126/science.1187191}{{Science}}
  \href{http://dx.doi.org/10.1126/science.1187191}{{\textbf{329}(5987), 69 LP
  }}\href{http://dx.doi.org/10.1126/science.1187191}{{ (2010)}}.

\bibitem{Gerlich2012}
D.~Gerlich, P.~Jusko, {\v{S}}.~Rou{\v{c}}ka, {et~al.}
\newblock \enquote{{Ion trap studies of H$^-$ + H $\rightarrow$ H$_2$ + e$^-$
  Between 10 and 135 K}}.
\newblock \href{http://dx.doi.org/10.1088/0004-637X/749/1/22}{{ApJ}}
  \href{http://dx.doi.org/10.1088/0004-637X/749/1/22}{{\textbf{749}(1)}}\href{http://dx.doi.org/10.1088/0004-637X/749/1/22}{{
  (2012)}}.

\bibitem{Otto2008}
R.~Otto, J.~Mikosch, S.~Trippel, {et~al.}
\newblock \enquote{{Nonstandard behavior of a negative ion reaction at very low
  temperatures}}.
\newblock \href{http://dx.doi.org/10.1103/PhysRevLett.101.063201}{{Phys. Rev.
  Lett.}}
  \href{http://dx.doi.org/10.1103/PhysRevLett.101.063201}{{\textbf{101}(6),
  1}}\href{http://dx.doi.org/10.1103/PhysRevLett.101.063201}{{ (2008)}}.

\bibitem{Gianturco2019}
F.~A. Gianturco, E.~Yurtsever, M.~Satta, and R.~Wester.
\newblock \enquote{{Modeling Ionic Reactions at Interstellar Temperatures: The
  Case of NH$_2^-$+ H$_2$ $\rightarrow$ NH$_3$ + H$^-$}}.
\newblock \href{http://dx.doi.org/10.1021/acs.jpca.9b07317}{{J. Phys. Chem. A}}
  \href{http://dx.doi.org/10.1021/acs.jpca.9b07317}{{\textbf{123}(46),
  9905}}\href{http://dx.doi.org/10.1021/acs.jpca.9b07317}{{ (2019)}}.

\bibitem{Keller2015}
E.~Jordan, G.~Cerchiari, S.~Fritzsche, and A.~Kellerbauer.
\newblock \enquote{{High-Resolution Spectroscopy on the Laser-Cooling Candidate
  La$^-$}}.
\newblock \href{http://dx.doi.org/10.1103/PhysRevLett.115.113001}{{Phys. Rev.
  Lett.}}
  \href{http://dx.doi.org/10.1103/PhysRevLett.115.113001}{{\textbf{115}(11),
  1}}\href{http://dx.doi.org/10.1103/PhysRevLett.115.113001}{{ (2015)}}.

\bibitem{Gerber2018}
S.~Gerber, J.~Fesel, M.~Doser, and D.~Comparat.
\newblock \enquote{{Photodetachment and Doppler laser cooling of anionic
  molecules}}.
\newblock \href{http://dx.doi.org/10.1088/1367-2630/aaa951}{{New J. Phys.}}
  \href{http://dx.doi.org/10.1088/1367-2630/aaa951}{{\textbf{20}(2),
  023024}}\href{http://dx.doi.org/10.1088/1367-2630/aaa951}{{ (2018)}}.

\bibitem{Wang2017}
M.-j. Wan, D.-h. Huang, Y.~Yu, and Y.-g. Zhang.
\newblock \enquote{Laser cooling of the OH$^-$ molecular anion in a theoretical
  investigation}.
\newblock \href{http://dx.doi.org/10.1039/C7CP04393G}{{Phys. Chem. Chem.
  Phys.}} \href{http://dx.doi.org/10.1039/C7CP04393G}{{\textbf{19},
  27360}}\href{http://dx.doi.org/10.1039/C7CP04393G}{{ (2017)}}.

\bibitem{Andresen2010}
G.~B. Andresen, M.~D. Ashkezari, M.~Baquero-Ruiz, {et~al.} (ALPHA
  Collaboration).
\newblock \enquote{{Evaporative Cooling of Antiprotons to Cryogenic
  Temperatures}}.
\newblock \href{http://dx.doi.org/10.1103/PhysRevLett.105.013003}{{Phys. Rev.
  Lett.}}
  \href{http://dx.doi.org/10.1103/PhysRevLett.105.013003}{{\textbf{105},
  013003}}\href{http://dx.doi.org/10.1103/PhysRevLett.105.013003}{{ (2010)}}.

\bibitem{Hlavenka2009a}
P.~Hlavenka, R.~Otto, S.~Trippel, {et~al.}
\newblock \enquote{{Absolute photodetachment cross section measurements of the
  O$^-$ and OH$^-$ anion}}.
\newblock \href{http://dx.doi.org/10.1063/1.3080809}{{J. Chem. Phys.}}
  \href{http://dx.doi.org/10.1063/1.3080809}{{\textbf{130}(6)}}\href{http://dx.doi.org/10.1063/1.3080809}{{
  (2009)}}.

\bibitem{Spitzer}
L.~Spitzer {{{Physics of fully ionized gases}}} 2nd ed.  (Interscience Publ.,
  New York, 1967).

\bibitem{Ichimaru1986}
S.~Ichimaru and S.~Tanaka.
\newblock \enquote{{Generalized viscoelastic theory of the glass transition for
  strongly coupled, classical, one-component plasmas}}.
\newblock \href{http://dx.doi.org/10.1103/PhysRevLett.56.2815}{{Phys. Rev.
  Lett.}}
  \href{http://dx.doi.org/10.1103/PhysRevLett.56.2815}{{\textbf{56}(26),
  2815}}\href{http://dx.doi.org/10.1103/PhysRevLett.56.2815}{{ (1986)}}.

\bibitem{Gilbert}
S.~L. Gilbert, J.~J. Bollinger, and D.~J. Wineland.
\newblock \enquote{{Shell-structure phase of magnetically confined strongly
  coupled plasmas}}.
\newblock \href{http://dx.doi.org/10.1103/PhysRevLett.60.2022}{{Phys. Rev.
  Lett.}}
  \href{http://dx.doi.org/10.1103/PhysRevLett.60.2022}{{\textbf{60}(20),
  2022}}\href{http://dx.doi.org/10.1103/PhysRevLett.60.2022}{{ (1988)}}.

\bibitem{Lauderdale1983}
J.~G. Lauderdale, C.~W. McCurdy, and A.~U. Hazi.
\newblock \enquote{Conversion of bound states to resonances with changing
  internuclear distance in molecular anions}.
\newblock \href{http://dx.doi.org/10.1063/1.446068}{{J. Chem. Phys.}}
  \href{http://dx.doi.org/10.1063/1.446068}{{\textbf{79}(5),
  2200}}\href{http://dx.doi.org/10.1063/1.446068}{{ (1983)}}.

\bibitem{Sanche1990}
L.~Sanche.
\newblock \enquote{Low-energy electron scattering from molecules on surfaces}.
\newblock \href{http://dx.doi.org/10.1088/0953-4075/23/10/005}{{J. Phys. B: At.
  Mol. Opt. Phys.}}
  \href{http://dx.doi.org/10.1088/0953-4075/23/10/005}{{\textbf{23}(10),
  1597}}\href{http://dx.doi.org/10.1088/0953-4075/23/10/005}{{ (1990)}}.

\bibitem{Engel2019}
F.~Engel, T.~Dieterle, F.~Hummel, {et~al.}
\newblock \enquote{Precision Spectroscopy of Negative-Ion Resonances in
  Ultralong-Range Rydberg Molecules}.
\newblock \href{http://dx.doi.org/10.1103/PhysRevLett.123.073003}{{Phys. Rev.
  Lett.}}
  \href{http://dx.doi.org/10.1103/PhysRevLett.123.073003}{{\textbf{123},
  073003}}\href{http://dx.doi.org/10.1103/PhysRevLett.123.073003}{{ (2019)}}.

\bibitem{Weckesser2021}
P.~Weckesser, F.~Thielemann, D.~Wiater, {et~al.}
\newblock \enquote{{Observation of Feshbach resonances between a single ion and
  ultracold atoms}}.
\newblock \href{http://dx.doi.org/10.1038/s41586-021-04112-y}{{Nature}}
  \href{http://dx.doi.org/10.1038/s41586-021-04112-y}{{\textbf{600}(7889),
  429}}\href{http://dx.doi.org/10.1038/s41586-021-04112-y}{{ (2021)}}.

\bibitem{Kas2019}
M.~Kas, J.~Loreau, J.~Li{\'{e}}vin, and N.~Vaeck.
\newblock \enquote{{Reactivity of Hydrated Hydroxide Anion Clusters with H and
  Rb: An ab Initio Study}}.
\newblock \href{http://dx.doi.org/10.1021/acs.jpca.9b05971}{{J. Phys. Chem. A}}
  \href{http://dx.doi.org/10.1021/acs.jpca.9b05971}{{\textbf{123}(41),
  8893}}\href{http://dx.doi.org/10.1021/acs.jpca.9b05971}{{ (2019)}}.

\bibitem{Hassan2022a}
S.~Z. Hassan, J.~Tauch, M.~Kas, {et~al.}
\newblock \enquote{{Associative detachment in anion-atom reactions involving a
  dipole-bound electron}}.
\newblock \href{http://dx.doi.org/10.1038/s41467-022-28382-w}{{Nat. Commun.}}
  \href{http://dx.doi.org/10.1038/s41467-022-28382-w}{{\textbf{13}(1),
  818}}\href{http://dx.doi.org/10.1038/s41467-022-28382-w}{{ (2022)}}.

\bibitem{Hassan2022b}
S.~Z. Hassan, J.~Tauch, M.~Kas, {et~al.}
\newblock \enquote{{Quantum state-dependent anion–neutral detachment
  processes}}.
\newblock \href{http://dx.doi.org/10.1063/5.0082734}{{J. Chem. Phys.}}
  \href{http://dx.doi.org/10.1063/5.0082734}{{\textbf{156}(9),
  094304}}\href{http://dx.doi.org/10.1063/5.0082734}{{ (2022)}}.

\bibitem{Thermometry}
M.~N{\"{o}}tzold, S.~Z. Hassan, J.~Tauch, {et~al.}
\newblock \enquote{{Thermometry in a Multipole Ion Trap}}.
\newblock \href{http://dx.doi.org/10.3390/app10155264}{{Appl. Sci.}}
  \href{http://dx.doi.org/10.3390/app10155264}{{\textbf{10}(15),
  5264}}\href{http://dx.doi.org/10.3390/app10155264}{{ (2020)}}.

\bibitem{Trippel2006}
S.~Trippel, J.~Mikosch, R.~Berhane, {et~al.}
\newblock \enquote{{Photodetachment of cold OH- in a multipole ion trap}}.
\newblock \href{http://dx.doi.org/10.1103/PhysRevLett.97.193003}{{Phys. Rev.
  Lett.}}
  \href{http://dx.doi.org/10.1103/PhysRevLett.97.193003}{{\textbf{97}(19),
  1}}\href{http://dx.doi.org/10.1103/PhysRevLett.97.193003}{{ (2006)}}.

\bibitem{Smith1997}
J.~R. Smith, J.~B. Kim, and W.~C. Lineberger.
\newblock \enquote{{High-resolution threshold photodetachment spectroscopy of
  OH$^-$}}.
\newblock \href{http://dx.doi.org/10.1103/PhysRevA.55.2036}{{Phys. Rev. A}}
  \href{http://dx.doi.org/10.1103/PhysRevA.55.2036}{{\textbf{55}(3),
  2036}}\href{http://dx.doi.org/10.1103/PhysRevA.55.2036}{{ (1997)}}.

\end{thebibliography}
\end{document}